\documentclass{elsart}
\usepackage{graphicx}
\usepackage{amssymb}

\begin{document}
\begin{frontmatter}

\title{Universal features of polymer shapes in crowded environment}

 \author[label1]{V.\  Blavatska\corauthref{cor1}}
 \ead{viktoria@icmp.lviv.ua}
 \corauth[cor1]{Corresponding author.}
 \author[label2,label3]{C.\ von Ferber}
\ead{c.vonferber@coventry.ac.uk}
 \author[label1,label4]{Yu.\ Holovatch}
 \ead{hol@icmp.lviv.ua}

\address[label1]{Institute for Condensed Matter Physics
 of the National Academy of Sciences of Ukraine,
                  UA-79011 Lviv, Ukraine}
                 \address[label2]{Applied Mathematics Research Centre, Coventry University,
 CV1 5FB Coventry, UK}
 \address[label3]{Physicalishes Institut, Universit\"at Freiburg,
                D-79104 Freiburg, Germany}
 \address[label4]{Institut f\"ur Theoretische Physik, Johannes Kepler
 Universit\"at Linz, A-4040 Linz, Austria}

\begin{abstract}
We study the universal characteristics of the shape of a polymer chain in an
environment with correlated structural obstacles,
 applying the field-theoretical renormalization group approach. Our results qualitatively
 indicate an increase of the asymmetry of the polymer shape in
crowded environment comparing with the pure solution case.
\end{abstract}

\end{frontmatter}


\section{Introduction} \label{I}

\begin{figure}[h!]
\begin{center}
\includegraphics[width=10cm]{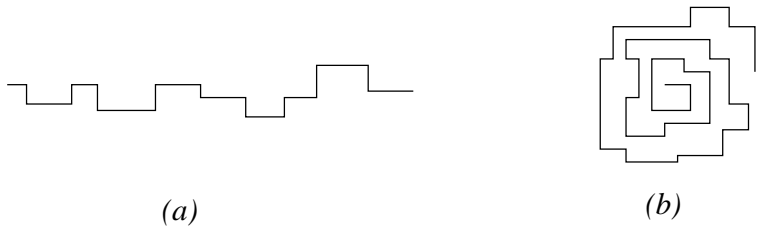}
\end{center}
\caption{\label{a}Schematic
presentation of a polymer configurations with aspericity ratio $\hat{A_d}$ value
close to one (a) and close to zero (b).
 }
\end{figure}

In studying the transport properties of polymer fluids, an important role is played by the shape characteristics
 of a single polymer chain configuration. It is established \cite{Kramers46,Solc71}, that the typical polymer chain realization does
 not possess spherical symmetry. A quantity, which characterizes the asymmetrical shape of a polymer chain, is the
asphericity ratio $\hat{A_d}$ of the chain configuration.  $\hat{A_d}$ attains a maximal value of one for a completely stretched,
 rod-like configuration,
and equals zero for the spherical form (see Fig. \ref{a}), thus obeying the inequality: $0\leq \hat{A_d} \leq 1$.
In the limit of long chains this quantity appears to be universal and depends on space dimension $d$ only: $\hat{A_d}>1/2$ at $d<4$, $\hat{A_d}=1/2$ at $d\geq 4$.
 The size measure of a flexible polymer chain is usually defined by either the mean-squared end-to-end distance $R_e$
 or  radius of gyration $R_G$.
The ratio of these quantities, the so-called size ratio $g\equiv\langle R_e^2 \rangle/\langle R_G^2 \rangle$ again is an
 universal quantity
 ($ g>6$  for $d<4$, $ g=6$  for $d\geq 4$).
The study of these universal quantities characterizing the polymer shape is a subject of a
great interest \cite{Bishop85,Rudnick86,Aronovitz86,Diel89}.

A current problem in a polymer physics is the influence of structural obstacles (impurities)
in the environment on the universal properties of macromolecules.
Such a disordered (crowded) environment can be found, in particular, in a biological cell,
composed of many different kinds of biochemical species, which may occupy
large fraction of the total volume \cite{Goodesel91}.
In the language of lattice models of polymers, the crowded environment with structural obstacles can be considered as a
disordered lattice, where a given fraction of randomly chosen sites are to be avoided by the polymer chain.
It has been proven both analytically \cite{Kim} and numerically \cite{Lee}, that
weak uncorrelated disorder that corresponds to the point-like randomly
distributed obstacles with concentration far from the percolation
threshold does not change the universality class of polymers.
In the present study, we address a model, where the structural obstacles of the environment are spatially
correlated on a mesoscopic scale \cite{Sahimi95}.
Following Ref. \cite{Weinrib83}, this case can be described by assuming the defects to be correlated on large distances $r$
according to a power law with a pair correlation function $
h(r)\sim r^{-a}. $ For $a<d$ such a correlation function describes defects extended in space
(see \cite{Weinrib83,Blavatska01a} for further details).
The impact of long-range-correlated disorder on the scaling of single polymer chains has been analyzed in
 previous works \cite{Blavatska01a} by means of the
 field-theoretical renormalization group (RG) approach.
 The intriguing question of the how the shape characteristics of a
 flexible chain are influenced by presence of such a disordered medium remains,
 however, still unresolved, and is the subject of the present study.

  The layout of the paper is as follows. In the next section,
 we describe the model of the flexible polymer chain in a crowded environment,
 develop its field-theoretical formulation and give a brief description of
  the field-theoretical RG approach. The results of
 its application to the present model and finally some conclusions and an outlook are given in the following sections.

\section{Field-theoretical description of the model}

Let $\vec{R}_n=\{x_n^{1},\ldots,x_n^d\}$ be the position vector of the $n$th monomer of the polymer chain ($n=1,\ldots,N$).
 The shape of a specific conformation of the chain can be  characterized \cite{Solc71} in terms of the gyration tensor $\bf{Q}$
with components:
\begin{equation}
Q_{ij}=\frac{1}{2N^2}\sum_{n,m=1}^N(x_n^i-x_m^i)(x_n^j-x_m^j),\,\,\,i,j=1,\ldots,d.
\label{mom}
\end{equation}
For the  averaged radius of gyration $R_G$ one thus has:
\begin{equation}
\langle R_G \rangle =\frac{1}{2N^2}\left\langle \sum_{n,m=1}^N
|\vec{R}_n-\vec{R}_m|^2 \right\rangle =  \left \langle \sum_{i=1}^d
Q_{ii} \right \rangle = \langle {\rm Tr}\,\bf{Q} \rangle. \label{rg}
\end{equation}
Here and below, $\langle \ldots \rangle$ denotes averaging over all
configurations of the polymer chain. The spread in eigenvalues
$\lambda_i$ of the gyration tensor measures the asymmetry of a given configuration.
In Ref.  \cite{Aronovitz86} it was proposed to characterize the
shape of macromolecule by ratios of rotationally invariant polynomials in the
components of $\bf{Q}$. Let us define the asphericity ratio
$\hat{A_d}$ as the quotient of two averaged quantities
\cite{Aronovitz86}:
\begin{equation}
\hat{A_d} =\frac{1}{d(d-1)} \sum_{i=1}^d\frac{\langle(\lambda_{i}-{\overline{\lambda}})^2\rangle}{\langle({\overline{\lambda}})^2\rangle}=\frac{d}{d-1}\frac{\langle{\rm {Tr}}\,{\bf{{\hat{Q}}}}^2\rangle}{\langle(\rm{Tr}\,{\bf{Q}})^2\rangle}. \label{add}
\end{equation}
Here, ${\overline{\lambda}}\equiv {\rm Tr}\, {\bf{Q}}/d$
is the mean eigenvalue of gyration tensor for a given chain configuration and ${\bf{\hat{Q}}}\equiv
{\bf{Q}}-{\overline{\lambda}}{\bf{I}}$ with the unity matrix $\bf{I}$.
$\hat{A_d}$ equals zero when all configurations are spherical (eigenvalues of each given configuration are equal) and takes
the maximal value of one when all configurations are rod-like  (all the eigenvalues equal zero except one).
Besides the ratio of averages one may also calculate the so-called
mean asphericity, given by an average of the ratio \cite{Jagod}. While ref. \cite{Jagod}
has shown some slight quantitative deviations between the two
approaches it confirmed that the qualitative dependence with respect
to other parameters, e.g. on the architecture of the molecule is the
same in each case.

Passing from the discrete model to a continuous polymer chain description  \cite{Edwards65} allows
  to derive a field-theoretical formulation of the asymptotic shape characteristics of a single polymer chain in an
  environment with correlated structural obstacles.
Applying the replica method in order to average the free energy over
different configurations of the randomly distributed and fixed
obstacles leads to an $m$-component field theory with a Lagrangian
$L_{{\rm {Dis}}}$
 \cite{Blavatska01a}:
\begin{eqnarray}
 L_{{\rm {Dis}}}&=& \frac{1}{2} \sum_{\alpha=1}^{n}
\int{\rm d}^d x [(\hat\mu_0^2|\vec{\phi}_{\alpha}(x)|^2+
|\nabla\vec{\phi}_{\alpha}(x)|^2)
+\frac{u_0}{4!}
(\vec{\phi}_{\alpha}^2(x))^2]
\nonumber\\
&&-\sum_{\alpha,\beta=1}^{n}
\int{\rm d}^dx{\rm d}^dy h(|x-y|)
\vec{\phi}_{\alpha}^2(x)\vec{\phi}_{\beta}^2(y). \label{h}
\end{eqnarray}
Here each replica  $\vec{\phi}_{\alpha}$ is an $m$-component vector
field
$\vec{\phi}_{\alpha}=(\phi_{\alpha}^1,\ldots,\phi_{\alpha}^m)$,
$\hat\mu_0$ and $u_0$ are bare mass and coupling, the coupling of
the replicas are described by the correlation function  $h(r)\sim
r^{-a}$ and both the polymer ($m\to 0$) and the replica ($n\to 0$)
limits  are implied. For small $k$, the Fourier-transform $\tilde
h(k)$ of $h(r)$ behaves as: $ \tilde h(k)\sim v_0+w_0|k|^{a-d}. $
Taking this into account, rewriting Eq.~(\ref{h}) in the momentum
space, and recalling special symmetry peculiarities of
 (\ref{h}) that appear for $m$, $n\to 0$ \cite{Blavatska01a},  a theory with
two bare couplings $u_0$, $w_0$ results.
Note that for $a\geq d$ the $w_0$-term is irrelevant in
the RG sense and one restores the pure case (absence of structural defects).

As shown in Ref. \cite{Aronovitz86}, computing the shape parameters of
long polymer chain can be reduced to computing the critical amplitudes and exponents of the corresponding field-theoretical model.
In order to extract the scaling behavior of the model (\ref{h})
we apply the RG method \cite{rgbooks}
to  get the Green's functions $G_R^{(L,N)}$ renormalized at non-zero mass and zero external momenta.
The change of couplings $u_0, w_0$ under renormalization defines a flow in the parametric space, governed by
corresponding $\beta$-functions: $
\beta_u(u,w)=\frac{\partial u}{\partial \ln
 \ell}\Big|_0,\,\,\beta_w(u,w)=\frac{\partial w}{\partial \ln
 \ell}\Big|_0$,
where $l$ is the rescaling factor, and $\Big|_0$   stands for
 evaluation at fixed bare parameters.
 The fixed points (FPs)
$u^*,w^*$ of the RG transformation are given by common zero of $\beta$-functions.
The stable FP corresponds to the
critical point of the system.

Following Ref. \cite{Aronovitz86}, the averaged moments of
the gyration tensor $\bf{Q}$ needed to determine the polymer shape
characteristics  (\ref{rg}) and (\ref{add})  can be expressed in terms of the
connected Green's function. In
particular:
\begin{eqnarray}
&&\langle Q_{ij}\rangle=-\frac{1}{2}\left( \frac{DT}{2\bar{X}}\right)^{2\nu}\frac{\Gamma(\gamma)}{\Gamma(\gamma+2\nu+2)}\frac{G_{ij}}{G_R^{(2)}(0,0,\{\lambda^*\})},
\label{qq}\\
&&\langle Q_{ij}Q_{kl}\rangle=-\frac{1}{4}\left( \frac{DT}{2\bar{X}}\right)^{4\nu}\frac{\Gamma(\gamma)}{\Gamma(\gamma+4\nu+4)}\frac{ G_{ij|kl}}  {G_R^{(2)}(0,0,\{\lambda^*\})},
\label{qqqq}
\end{eqnarray}
where $D, T, \bar{X}$ are non-universal quantities which will drop out in the final expressions,
$\Gamma$ is the Euler gamma-function,  $\nu$ and $\gamma$ are the
critical exponents of model (\ref{h}), and the following notations are used:
\begin{eqnarray}
&&G_{ij}\equiv \left(\frac{\partial}{\partial q^{i}}\frac{\partial}{\partial q^{j}}G_R^{(2,2)}(0,0;{\bf q},-{\bf q};\{\lambda^*\})\right)\Big{|}_{\{{\bf q}\}=0}, \\
&&G_{ij|kl}= \left(\frac{\partial}{\partial q^{i}_1}\frac{\partial}{\partial q^{j}_1}\frac{\partial}{\partial q^{k}_2}\frac{\partial}{\partial q^{l}_2}G_R^{(2,4)}(0,0;{\bf q}_1,-{\bf q}_1,{\bf q}_2,-{\bf q}_2;\{\lambda^*\})\right)\Big{|}_{\{{\bf q}\}=0}.
\end{eqnarray}
 Here $G_R^{(2,2)}(0,0;{\bf q},-{\bf q};\{\lambda^*\})$ and
 $G_R^{(2,4)}(0,0;{\bf q}_1,-{\bf q}_1,{\bf q}_2,-{\bf q}_2;\{\lambda^*\})$ are the fixed point values of the
 renormalized two-point connected Green's
 functions with two and four $\phi^2$-insertions, the symbol $\Big{|}_{\{{\bf q}\}=0}$ indicates that the corresponding
 expressions are to be taken at all external momenta $\{{\bf q}\}$ equal to zero.

The isotropy of the model implies, in particular, that $\langle {\rm Tr\, {\bf {Q}}}
\rangle =d\langle Q_{xx} \rangle$, so that:
\begin{equation}
\langle R_G \rangle = d \langle Q_{xx} \rangle.
\end{equation}
 For the mean-squared end-to-end distance $\langle R_e^2 \rangle$ one has \cite{Aronovitz86}:
\begin{equation}
\langle R_e^2 \rangle=-\left( \frac{DT}{2\bar{X}}\right)^{2\nu}
\frac{\Gamma(\gamma)}{\Gamma(\gamma+2\nu)}\frac{\left(\nabla_{\bf{k}}^2G_R^{(2)}({\bf k},-{\bf k},
\{\lambda^*\})\right)|_{\bf{k}=0}}{G_R^{(2)}(0,0,\{\lambda^*\})}.
\end{equation}
where $\nabla_{\bf{k}}^2$ means differentiation over components of external moment $\bf{k}$.
The not-universal quantities cancel when  the ratio $ g=\langle R_e^2\rangle /\langle R_G^2 \rangle$ is considered:
\begin{equation}
\label{ratio0}
  g = \frac{2\Gamma(\gamma+2\nu+2)}{\Gamma(\gamma+2\nu)}
 \frac{\left(\nabla_{\bf{k}}^2G_R^{(2)}({\bf k},-{\bf k},\{\lambda^*\})\right)|_{\bf{k}=0}}
 {\left(\nabla_{\bf{q}}^2G_R^{(2,4)}(0,0;{\bf q},-{\bf q};\{\lambda^*\})\right)|_{{\bf q}=0}}
\end{equation}
and thus the size ratio is  universal quantity.
The asphericity  ratio $\hat{A_d} $ can be expressed in terms of the averaged moments of gyration tensor (\ref{qq}), (\ref{qqqq}) as follows:
 \begin{equation}
\hat{A_d}  =\frac{\langle Q_{xx}^2\rangle+d\langle Q_{xy}^2\rangle-\langle Q_{xx}Q_{yy}\rangle}
{\langle Q_{xx}^2\rangle+d(d-1)\langle Q_{xx}Q_{yy}\rangle}. \label{adfinal}
\end{equation}
One can again easily convince oneself, that all  non-universal quantities in Eqs. (\ref{qq}), (\ref{qqqq})
cancel when calculating (\ref{adfinal}), and $\hat{A_d} $ is a universal quantity.


\section{Results}
\label{II}

To obtain the qualitative
characteristics of the critical behavior of the model, we exploit a double expansion in the
parameters $\varepsilon=4-d$ and $\delta=4-a$, assuming them to be of the same order of magnitude.
Within this approach it was shown \cite{Blavatska01a}, that a single polymer chain in a solvent in an environment with
long-range-correlated structural obstacles belongs to a universality class different from the case of a pure solvent.
In the  field-theoretical renormalization group description, this is reflected by the appearance of a new non-trivial stable
long-range-correlated (LR) fixed point besides the usual (pure) one.
The coordinates of these FPs and their regions of stability read \cite{Blavatska01a}:
\begin{eqnarray}
\left \{ \begin{array}{ll}{\rm pure\,\, FP}: u^*=\frac{3\varepsilon}{4},\,w^*=0 & \,\,\,\,{\rm at}\,\,\delta<\varepsilon/2, \\
{\rm LR\,\,\,\,FP}: u^*=\frac{3\delta^2}{2(\delta-\varepsilon)},\, w^*=\frac{3\delta(\varepsilon-2\delta)}{2(\varepsilon-\delta)}
 &\,\,\,{\rm at}\,\,
\varepsilon/2 <\delta<\varepsilon.  \end{array}
\right.
\label{LR}
\end{eqnarray}

To estimate the size ratio (\ref{ratio0}) for the case of a polymer
in long-range-correlated disorder, we calculate the  function
$G_R^{(2,2)}(0,0;{\bf q},-{\bf q};u,w)$ with two insertions
$\phi^2({\bf{q}})$, $\phi^2(-{\bf{q}})$. In the first order of the
$\varepsilon=4-d$, $\delta=4-a$-expansion  we find:
\begin{eqnarray}
&&G_R^{(2,2)}(0,0;{\bf q},-{\bf q};u,w)=\frac{2}{q^2+1}-\frac{4}{3} \frac{1}{q^2+1}[u I_1-w J_1]
-\frac{2}{3}[ u I_2-w J_2]\nonumber\\
&&+ \frac{4}{3}\frac{1}{q^2+1}[ u I_0-w J_0].
\end{eqnarray}
Here $I_i$, $J_i$ are given by  the following one-loop integrals:
\begin{eqnarray}&&I_0=\int\frac{{\rm d}\vec{p}}{(p^2+1)^2},\,\,\,\,\,\,\,\,\,\,\,\,\,\,\,\,
J_0=\int\frac{{\rm d}\vec{p}\,|p|^{a-d}}{(p^2+1)^2}\label{i0}\\
&& I_1=\int\frac{{\rm d}\vec{p}}{(p^2+1)((p+q)^2+1)},\,\,\,\,J_1=\int\frac{{\rm d}\vec{p}\,|p|^{a-d}}{(p^2+1)((p+q)^2+1)},\label{i1}\\
&& I_2=\int\frac{{\rm d}\vec{p}}{(p^2+1)^2((p+q)^2+1)},\,\,\,\,J_2=\int\frac{{\rm d}\vec{p}\,|p|^{a-d}}{(p^2+1)^2((p+q)^2+1)}.\label{i2}
\end{eqnarray}
To evaluate $g$ according to Eq. (\ref{ratio0}), we perform the derivation of  $G_R^{(2,2)}$  with respect to the  components
of the vector ${\bf q}$ and expand the loop integrals
in $\varepsilon$ and $\delta$.
Further  inserting the FP values (\ref{LR}) we finally receive:
\begin{equation}
 g = \left \{ \begin{array}{ll}  g^{{\rm pure}}= 6+\frac{\varepsilon}{16},\; & \delta<\varepsilon/2, \\
 g^{{\rm LR}} = 6+\frac{\delta}{8},\; &
\varepsilon/2 <\delta<\varepsilon.  \end{array}
\right.\label{ratio}
\end{equation}
Let us qualitatively estimate the change in the size ratio $ g$, caused by presence of the structural obstacles,
 in three dimensions. Substituting directly $\varepsilon=1$ into the first line of (\ref{ratio}), we have for the
 polymer chain in a pure solvent:
 $ g^{{\rm pure}}~\simeq 6.06. $  Let us recall, that the influence of the long-range-correlated disorder is relevant
 for $a\leq d$ ($\delta\geq\varepsilon$) (see e.g. explanation after Eq. (\ref{h})).
 Estimates of $ g^{{\rm LR}}$ can be evaluated by direct substitution of continuously changing parameter
 $\delta$ into the second line of Eq. (\ref{ratio}).
 One concludes, that increasing the parameter $\delta$ (which corresponds to an increase of strength
 of  disorder) leads to corresponding increase of the ratio of
 the end-to-end to the gyration radii $g$.

To compute the averaged asphericity ratio using (\ref{adfinal}), we calculate the
 function $G_R^{(2,4)}(0,0;{\bf q}_1,-{\bf q}_1,{\bf q}_2,-{\bf q}_2;u,w)$
 with four insertions $\phi^2({\bf{q}}_1)$, $\phi^2(-{\bf{q}}_1)$, $\phi^2({\bf{q}}_2)$, $\phi^2(-{\bf{q}}_2)$.
 The resulting expansion is too cumbersome to be presented here and will be given elsewhere \cite{new}.
Performing the derivatives with respect to the components of the vectors $\bf{q}_1$, $\bf{q}_2$ we
find:
\begin{eqnarray}
G_{xx}=576+\frac{4028}{15}(u-w),
\nonumber\\
G_{xx|yy}=320+\frac{436}{3}(u-w),
\nonumber\\
G_{xy|xy}=128+\frac{308}{5}(u-w).\label{di}
\end{eqnarray}
Inserting the FP values  (\ref{LR}) into Eqs. (\ref{di}) and recalling the definitions
(\ref{qq}) and (\ref{qqqq}) the result is:
\begin{equation}
\hat{A_d}  =\left \{ \begin{array}{ll} \hat{A_d}^{{\rm pure}}= \frac{1}{2}+\frac{15}{512}\,\varepsilon,\; & \delta<\varepsilon/2, \\
\hat{A_d}^{{\rm LR}}=\frac{1}{2}+\frac{1}{48}\,\varepsilon+\frac{13}{768} \,\delta,\; &
\varepsilon/2 <\delta<\varepsilon.  \end{array}
\right.\label{adall}
\end{equation}
Again, let us qualitatively estimate the change in $\hat A_d $  caused by
the presence of structural obstacles in three dimensions. Substituting directly $\varepsilon=1$ into the
first line of (\ref{adall}), we have for the pure case: $ \hat{A_d}^{{\rm pure}}\simeq 0.53.
 $ Estimates of $\hat{A_d}^{{\rm LR}}$ can be obtained by direct
substitution of the continuously changing parameter $\delta$ into the second line of Eq. (\ref{adall}).
Increase of the strength of disorder correlations results in
increase of the asphericity  ratio of polymers in disorder.
This phenomenon can be intuitively understood if one recalls an impact
of the long-range-correlated disorder on the mean end-to-end distance
exponent $\nu$. Indeed, it has been shown in \cite{Blavatska01a}, that such a disorder leads to
an increase of $\nu$, and subsequently, to further swelling of the polymer chain.
Extended obstacles disfavor return trajectories and as a result the polymer chain
  becomes more elongated. This elongation then leads to an increase of the
asphericity ratio as predicted by Eq. (20).


\section{Conclusions and outlook}
\label{III}
The universal characteristics of the average shape of a polymer coil configurations in a
porous (crowded) environment with structural obstacles have been analyzed considering the special case,
when defects are correlated  at large distances $r$ according to the power law: $h(r)\sim r^{-a}$.
Applying the field-theoretical RG approach, we estimate the size
 ratio $ g=\langle R_e^2 \rangle/\langle R_G^2 \rangle$ and averaged asphericity ratio $\hat{A_d}$ up
 to the first
 order of a double $\varepsilon=4-d$, $\delta=4-a$ expansion. We have revealed, that
 the presence of long-range-correlated disorder leads to an increase of both $g$ and $\hat{A_d}$
 as compared to their values for a polymer chain in a pure solution. Moreover,
 the asphericity ratio $\hat{A_d}$ value was found to be closer
to the maximal value of one in presence of correlated obstacles.
Thus, we  conclude, that the presence of structural obstacles in an environment
makes the polymer coil configurations to be less spherical.
Let us note that the present results,
 obtained in the first order of an $\varepsilon$, $\delta$-expansion should serve as a qualitative
 estimate rather than an accurate numerical evaluation. The next step in our analysis will be to
 obtain the higher order expansions for the quantities of interest and to perform computer simulations
 in order to confirm these theoretical results by numerical values.
 Further details of our calculations as well as a generalization of the presented results to
  the case of polymers with complex topology
 will be given elsewhere \cite{new}.

\section*{Acknowledgements}
The support of the Alexander von Humboldt
Foundation and National Academy  of Sciences of Ukraine Committee for Young Scientists
 (V.B.), the Austrian FFWF project No.19583-N20 (Yu.H.) and an Applied  Research Fellowship of Coventry University (C.vF.)
 is greatly acknowledged.



\end{document}